\documentclass[reviewer,10pt,authoryear,fleqn]{elsarticle}
\usepackage[margin=1in]{geometry}
\usepackage{amsmath}
\usepackage{amssymb}
\usepackage{graphicx}
\usepackage{hyperref}

\journal{New Astronomy}

\begin{document}

\begin{frontmatter}
\title{Acoustic geometry obtained through the perturbation of the Bernoulli's constant}	
\author{Satadal Datta}
\ead{satadaldatta@hri.res.in}
\author{Md Arif Shaikh}
\ead{arifshaikh@hri.res.in}
\author{Tapas Kumar Das}
\ead{tapas@hri.res.in}

\address{Harish-Chandra Research Institute, HBNI, Chhatnag Road, Jhunsi, Allahabad 211019, India}

\begin{abstract}
For accretion onto astrophysical black holes, we demonstrate that linear perturbation of Bernoulli's constant defined for an inviscid irrotational adiabatic flow of perfect ideal fluid gives rise to phenomena
of analogue gravity.  The formulation of our work is done in the Newtonian framework and also within the General relativistic framework, i.e., considering a static spacetime background, as well.  The resulting
structure of the analogue acoustic metric is similar to the acoustic metric found in perturbing velocity potential and mass accretion rate. 
\end{abstract}

\end{frontmatter}

\section{Introduction}
Except for the supersonic stellar wind fed accretion (\cite{lamers1999introduction,Frank1985accretion,accretion-japan2008}), accretion flows onto astrophysical black holes are necessarily supersonic (\cite{Liang1980transonic}). For low angular momentum accretion with practically constant specific angular momentum, more than one sonic points may form in such flow and a stationary shock may join two such transonic solutions passing through two such sonic points (\cite{Liang1980transonic,Abramowicz1981,Muchotrzeb1982,Muchotrzeb1983,Fukue1983,Fukue1987,Lu1985,Lu1986,Muchotrzeb1986,Abramowicz1989,Abramowicz1990,CHAKRABARTI_physics_Reports,Kafatos1994,Yang1995,Pariev1996,Peitz1997,Caditz1998,TKD2002,TKD-Mitra-2003,Barai2004,Fukue2004a,Fukue2004b,abraham:causal,Das-bilic-dasgupta,TKD2012-Czerny}).  The formation of such shocks can be explained through time dependent numerical 
simulation works (\cite{Okuda2004,Okuda2007,Sukova2015JOP,Sukova2015MNRAS,Sukova2017}). Such post shock flow can manifest its properties through the characteristic black hole spectra and can help to understand the observational signature of the 
astrophysical black holes in the universe (\cite{DAS201581,Monika} and references therein). Such shocked multi-transonic flows are essentially barotropic, inviscid, irrotational transonic fluid flow under the influence of the strong gravitational field in presence of gravitational (black hole) event horizon. 

It is, however, to be noted that the characteristic black hole spectra for transonic accretion are usually computed for steady state flow, considering that such steady state is stationary. Time variability and various kind of fluctuations 
are, however, not very uncommon in large scale astrophysical flows. It is thus imperative to ensure that the steady state integral transonic accretion solutions are stable under perturbation. 

In recent years, much attention have been paid to study the analogue gravity phenomena in classical (non quantum), where for a  supersonic irrotational inviscid flow governed by a barotropic equation of state, the propagation of the linear acoustic perturbation (sound wave) within that fluid can be described by acoustic metric and a sonic spacetime can be formed embedded within such stationary background fluid flow (\cite{Unruh,Visser1998,Barcelo,Novello-visser}). Such sonic geometry contains an acoustic horizon from where Hawking like radiation may be produced. 

Study of such sonic geometries embedded within the transonic accretion flow can thus be very important to investigate certain novel features of such phenomena. Accreting black holes is the only system found in the universe where both type of horizons, gravitational as well as acoustic, can be formed, and the same fluid can pass through both type of horizons as well. Hence theoretically if one would like to compare the properties of these two types of horizons, accreting black holes may be considered as the best candidate to study the sonic geometry embedded within it. Also, in usual analogue models, the gravitational field does not play any role while formulating the corresponding sonic geometry. For accretion onto astrophysical black holes (for accretion onto any compact massive astrophysical objects in general), the gravity determines the dynamics of the fluid and hence the associated acoustic spacetime itself is influenced by the gravitational field. 

For purely classical analogue systems, the detailed analysis of the quantum Hawking like effects may not always be possible to study, however, the study of the acoustic surface gravity can have deep significance in such systems. The acoustic surface gravity itself is a rather crucial entity to understand the flow structure as well as the associated sonic metric, and can thus be studied independently without looking into the existence of any analogue radiation (of phonons) like thermal phenomena characterized by their very feeble temperature too impractical to detect through any present day experimental set up. In recent years, the role of the analogue surface gravity in studying the non negligible effects associated with the emergence of the stimulated Hawking effects has been highlighted by examining such effects through the modified dispersion relations. Such study
have been performed from the purely analytical point of view as well as within the experimental set up (\cite{Rousseaux2008,Rousseaux2010,Jannes2011,WeinfurtnerPRL2011,Leonhardt2012,Robertson2012})

The deviation of the Hawking like effects within a dispersive media, i.e., within the fluid under consideration, from the universal behaviour of the original Hawking effect, depends sensitively on the gradient of the background bulk stationary velocity, as it has recently been suggested. It is, however, important to note that such theory of the non universal feature of the Hawking radiation has been postulated essentially for the isothermal flow and hence the space gradient for the sound velocity has not been taken into account. Also, the exact numerical values corresponding to the velocity gradient has not been possible to found yet and has been approximated by making certain assumptions.

For stationary integral accretion solutions as discussed in aforementioned paragraphs, the values of the space gradient of {\it both} the dynamical flow velocity as well as the speed of propagation of the acoustic perturbation have been computed very accurately using numerical schemes. Hence it is obvious that if the accreting black hole system can be studied as a classical analogue system, the non universal features of the Hawking like effects can further be modified including certain novel features.

It is thus obvious that the accreting black hole systems, although may not provide any direct signature of the Hawking like temperature (analogue temperature arising out from the phonon radiation) can still be considered as a very important as well as a unique theoretical construct to study the analogue gravity effects. 

The study of the stability properties of the background integral accretion solutions can lead to the emergence of the analogue 
phenomena which we shall demonstrate in the present work. We shall linear perturb various accretion related physical variables and will show that such perturbation does not grow indefinitely and hence the steady state accretion flow is actually stable under linear perturbation. We shall not perform any non linear stability analysis in the present work. We shall also show that such perturbation analysis leads to the formation of the acoustic metric embedded within the fluid flow. 

In this connection, it is to be mentioned that one can make attempt to study the analogue phenomena for the primordial micro black holes as well for which the analogue as well as the original Hawking temperature will be significantly large and will perhaps be a measurable quantity. It is, however, to be noted that the kind of analogue system we are interested in, can be emerged automatically in presence of accretion only. The theory of accretion flows onto primordial black hole are not a very well understood phenomenon as of now. To study the analogue effects for such black hole systems, one first has to formulate a self consistent theory of accretion processes around micro black holes, which, in our opinion, is a rather involved task to accomplish, and is obviously beyond the scope of the present work. Hence in the present work, we concentrate 
on large astrophysical black holes to study the analogue effects.

Mass accretion rate is a quantity having a reasonable physical significance in accretion phenomena. Linear perturbation of mass accretion rate in sub-Keplerian disk accretion in non relativistic framework also behaves like a massless scalar field in curved space-time(\cite{Nag_role_of_flow_geometry}), i.e., analogue gravity also emerges when accretion rate is perturbed.

Several works have been done in general relativistic framework as well. Linear perturbation of velocity potential in curved space-time background shows analogue gravity effect(\cite{Bilic1999}). Similarly, linear perturbation of mass accretion rate in accretion of perfection fluid in curved space-time background also shows same effect(\cite{deepika-sph2015,deepika_ax_schwarzchild,Bilic1999}).

In this work, we've shown that linear perturbation of another quantity, the Bernoulli's constant which is the integral solution of the corresponding Euler equation, also produces similar acoustic geometry. The whole work is being done in the non-relativistic framework and relativistic framework as well. Accretion phenomena of adiabatic flow are chosen to illustrate the fact. Radial accretion having spherical symmetry as well as disk accretion having axial symmetry are considered.

The linear perturbation technique also has astrophysical significance. We get a wave equation of the linear perturbation of Bernoulli's constant which is similar to the massless scalar field Klein-Gordon equation in curved space-time geometry. The nature of the solution of this wave equation tells us whether the existing steady state solutions like steady state solution for Bondi accretion, are stable or not under such perturbation in the medium. We have done stability analysis for that and we have concluded that not only the integrals of motion play a crucial role to determine the dynamics of the accretion flow in steady state but also their linear perturbations govern the behaviour of all the dynamical and thermodynamic quantities in the time dependent problem within the perturbative framework.

In our work, the correspondence between a classical (non-quantum) 
	analogue model and the accretion processes onto astrophysical black holes
	has been established through the process of linear stability analysis of 
	stationary integral transonic 
	accretion solutions corresponding to the steady state 
	flow only. That means, only such accreting black hole systems have 
	been considered which are in steady state. The body of literature in accretion
	astrophysics, however, is huge and diverse. There are several steady state 
	flow models which may not be multi-transonic, and there are several excellent 
	works which deal with non steady hydrodynamic accretion (which may not 
	contain multiple sonic points or shocks) which may include
	various kind of time variabilities and instabilities, using complete 
	time-dependent numerical simulation (\cite{Hawley1984a,Hawley1984b,Kheyfets1990,Hawley1991,Yokosawa1995,Igumenshchev1996,Igumenshchev1997,Nobuta1999,Molteni1999,Stone1999,Caunt2001,DeVilliers2002,Proga2003,Gerardi2005,Moscibrodzka2008,Nagakura2008,Nagakura2009,Janiuk2009,Bambi2010a,Bambi2010b,Barai2011,Barai2012,Sukova2015MNRAS,Zhu2015,Narayan2016,Moscibrodzka2016,Sukova2017,Sadowki2017,Mach2018,Karkowski2018,Inayoshi2018,Fragile2018}). 
	We, however, did not concentrate on such approach. In the present paper,
	our main motivation was to explore how the analogue gravity phenomena can be 
	addressed through the linear stability analysis of steady-state solutions of
	hydrodynamic accretion.
\section{Acoustic gravity in non-relativistic framework}
 In non-relativistic frame work, fluid velocity is much less than light speed. The momentum conservation equations  and mass conservation equation for fluid is taken(\cite{clarke}) according to Newton's laws of dynamics. 
The continuity equation of fluid is given by
\begin{equation}
\frac{\partial \rho}{\partial t}+\vec{\nabla}.(\rho \vec{v})=0
\end{equation} 
where $\rho,\vec{v}$ are fluid density and velocity respectively.
Euler momentum equation for inviscid flow in an external field in general is given by
\begin{equation}
\frac{\partial \vec{v}}{\partial t}+\vec{v}.\vec{\nabla}\vec{v}=-\nabla\psi-\frac{\vec{\nabla p}}{\rho} 
\end{equation}
where the potential function of the external conservative field is $\psi$ and pressure at any point of the fluid is $ p $.\\
The flow is taken to be irrotational.
\begin{equation}
\vec{\nabla}\times\vec{v}=0
\end{equation}
Using irrotationality condition one can write Euler equation as
\begin{equation}
\frac{\partial \vec{v}}{\partial t}+\vec{\nabla}(\frac{1}{2}\vec{v}^2+\int \frac{dp}{\rho}+\psi)=0
\end{equation}
Bernoulli's constant, $\zeta$ is given by
\begin{equation}
\zeta=\frac{1}{2}\vec{v}^2+\int \frac{dp}{\rho}+\psi
\end{equation}
For steady irrotational flow it's a constant along the streamline. Adiabatic sound speed is given by
\begin{equation}
c_{s}^{2}=\frac{dp}{d\rho}=\frac{\gamma p}{\rho}
\end{equation}
where $\gamma$ is the specific ratio of the ideal gas.
We assume there is a stationary solution in general for the above equations, and we introduce linear perturbations in the fluid discussed in the next section.
\subsection{General procedure to obtain the acoustic metric}
Linear perturbation of fluid velocity and fluid pressure is introduced as 
\begin{align*}
\vec{v}(\vec{x},t)=\vec{v}_0(\vec{x})+\vec{v'}(\vec{x},t)
\end{align*}
\begin{align*}
\rho(\vec{x},t)=\rho_0(\vec{x})+\rho'(\vec{x},t)
\end{align*}
where $\rho_0(\vec{x})$, $\vec{v}_0(\vec{x})$ are the stationary solution of fluid density and velocity field and $\vec{v'}(\vec{x},t)$, $\rho'(\vec{x},t)$ are the introduced linear perturbation terms in the velocity and the density of the fluid.\\
As a result, the linear perturbation term in Bernoulli's constant is given by
\begin{equation}
\zeta'=\vec{v{_0}}.\vec{v'}+\frac{c_{s0}^2}{\rho_0}\rho'
\end{equation}
$c_{s0}$ is the stationary unperturbed sound speed. Continuity equation in terms of linear perturbation is given by
\begin{equation}
\frac{\partial\rho'}{\partial t}+\vec{\nabla}.(\rho'\vec{v_0}+\rho_0\vec{v'})=0
\end{equation}
Momentum equation in terms of linear perturbation is given by
\begin{equation}
\frac{\partial\vec{v'}}{\partial t}+\vec{\nabla}(\zeta')=0
\end{equation}
We have used equation (7) to find the above equation. Using equation (7) and equation (9) and taking another partial time derivative in equation (8) we get
\begin{equation}
\partial_{\mu}(f^{\mu\nu}(\vec{x})\partial_{\nu})\zeta'(\vec{x},t)=0
\end{equation}
where $f^{\mu\nu}(\vec{x})$ in Cartesian coordinate is given by
\begin{equation}
f^{\mu\nu}(\vec{x})=\frac{\rho_{0}}{c_{s0}^{2}}\begin{bmatrix}
-1 & \vdots & -v_{0}^{j} \\
\cdots&\cdots&\cdots\cdots \\
-v_{0}^{j}&\vdots & c_{s0}^{2}\delta^{ij}-v_{0}^{i}v_{0}^{j}
\end{bmatrix}
\end{equation}
where $i, j$ run over 1, 2, 3 representing three spatial dimensions. This $f^{\mu\nu}$ is exactly the same as $f^{\mu\nu}$ obtained when velocity potential is perturbed(\cite{Visser1998})\\
Massless scaler field equation in a spacetime background is given by
\begin{equation}
\square\varphi=\frac{1}{\sqrt{-g}}(\partial_{\mu}\sqrt{-g}g^{\mu\nu}\partial_{\nu})\varphi=0
\end{equation}
where $\rm \varphi$ is the scalar field, $g^{\mu\nu}$ is the background metric and $g$ is the determinant of the metric. Comparing equation (12) and equation (10) 
\begin{equation}
f^{\mu\nu}=\sqrt{-g}g^{\mu\nu}
\end{equation}
Immediately one can get 
\begin{equation}
det(f^{\mu\nu})=(\sqrt{-g})^4g^{-1}=g=-\frac{\rho_0^4}{c_{s0}^2}
\end{equation}
So g$^{\mu\nu}$ is given by
\begin{equation}
g^{\mu\nu}(\vec{x})=\frac{1}{\rho_{0}c_{s0}}\begin{bmatrix}
-1 & \vdots & -v_{0}^{j} \\
\cdots&\cdots&\cdots\cdots \\
-v_{0}^{j}&\vdots & c_{s0}^{2}\delta^{ij}-v_{0}^{i}v_{0}^{j}
\end{bmatrix}
\end{equation}
The acoustic metric is
\begin{equation}
g_{\mu\nu}(\vec{x})=\frac{\rho_{0}}{c_{s0}}\begin{bmatrix}
 -(c_{s0}^{2}-v_{0}^{2}) & \vdots & -v_{0}^{j} \\
\cdots&\cdots&\cdots\cdots \\
-v_{0}^{j}&\vdots &\delta_{ij}
\end{bmatrix}
\end{equation}
Acoustic metric interval can be expressed as
\begin{equation}
ds^{2}=\frac{\rho_{0}}{c_{s0}}\left[-(c_{s0}^{2}-v_{0}^{2})dt^{2}-2dt\vec{v_{0}}.d\vec{x}+d\vec{x}^{2} \right]
\end{equation}
 The same kind of analysis can be done for isothermal flow as well and the metric will be same except that the definition of sound speed will be different there, in the acoustic metric, sound speed will be appearing as a constant rather than a function of position vector.
 
The metric appearing in equation (16) has 3+1 dimension. It reduces to 1+1 dimension when symmetries in the flow is considered. The next section deals with some astrophysical accretion phenomenon having different kind of symmetries. 
\subsection{Spherically symmetric radial flow}
Bondi accretion(\cite{Bondi1952}) is spherically symmetric and radial. The continuity equation is given by
\begin{equation}
\frac{\partial \rho}{\partial t}+\frac{1}{r^2}\frac{\partial }{\partial r}(\rho vr^2)=0
\end{equation}
Euler momentum equation is given by
\begin{equation}
\frac{\partial v}{\partial t}+v\frac{\partial v}{\partial r}=-\frac{GM}{r^2}-\frac{1}{\rho}\frac{\partial p}{\partial r}
\end{equation}
where $M$ is the mass of the star and $G$ is gravitational constant. Bernoulli's constant is given by
\begin{equation}
\zeta=\frac{1}{2}v^2+\int \frac{dp}{\rho}-\frac{GM}{r}
\end{equation}
Introducing linear perturbation in adiabatic flow
\begin{align*}
v(r,t)=v_0(r)+v(r,t)'
\end{align*}
\begin{align*}
\rho(r,t)=\rho_0(r)+\rho(r,t)'
\end{align*}
Perturbation in Bernoulli's constant is given by 
\begin{equation}
\zeta'=v_0 v'+\frac{c_{s0}^2}{\rho_0}\rho'
\end{equation}
Now in the same way discussed earlier we find that the linear perturbation of Bernoulli's constant obeys massless scalar field equation in acoustic analogue of spacetime background.
\begin{equation}
\partial_{\mu}(f^{\mu\nu}(r)\partial_{\nu})\zeta'(r,t)=0
\end{equation}
where
\begin{equation}
f^{\mu\nu}(r)=\frac{\rho_{0}r^2}{c_{s0}^{2}}\begin{bmatrix}
-1  &~~~ -v_{0} \\
-v_{0}&~~~ c_{s0}^{2}-v_{0}^{2}
\end{bmatrix}
\end{equation}
$f_{\mu\nu}$ is taken as effective metric(\cite{Nag_role_of_flow_geometry}). Hence 2$\times$2 effective acoustic metric is given by
\begin{equation}
g_{\mu\nu}^{eff}(r)=\frac{1}{\rho_{0}r^2}\begin{bmatrix}
 -(c_{s0}^{2}-v_{0}^{2}) &~~~ -v_{0}\\
-v_{0} &~~~ 1
\end{bmatrix}
\end{equation}
Observation of the acoustic metric shows that acoustic horizon is produced.\\
Similarly, the same analysis can be done for isothermal flow as well.
\subsection{Axially symmetric sub-Keplarian disk geometries}
Low angular momentum axisymmetric blackhole accretion(\cite{Chakrabarti2001,Paczynski1982}) is a good candidate where analogue gravity emerges too(\cite{Nag_role_of_flow_geometry}). We don't need to consider viscosity in such weakly rotating sub-Keplarian flows due to low angular momentum of the infalling fluid. There are mainly three disk models for sub-Keplarian disk by categorising them with respect to disk height or thickness $H$. $H$ is taken to be constant in the simplest possible model, i.e., in uniform thickness disk model. In conical model(\cite{Chakrabarti2001}) the disk thickness $H$ is proportional to cylindrical radial distance from the accretor. In the most physical disk model, i.e., in Vertical equilibrium model(\cite{Paczynski1982,Chaudhury2006}), the disk height $H(r)$ is a function of cylindrical radial distance $r$ from the accretor such that there is no flow along $z$ direction considering the equatorial plane of the disk to be on the $X-Y$ plane. In this type of models, due to symmetry, the problem becomes effectively 1+1 dimensional. In the next section we consider the non-trivial disk model, i.e., vertical equilibrium disk model.
\subsection{Vertical equilibrium disk accretion}
Continuity equation in cylindrical polar coordinate in disk accretion having axial symmetry and having no net flow in z direction is given by
\begin{equation}
\frac{\partial \bar{\rho}(r,z)}{\partial t}+\frac{1}{r}\frac{\partial }{\partial r}(\bar{\rho}(r,z) \bar{v}(r,z)r)=0
\end{equation}
where $\bar{\rho}(r,z)$ and $\bar{v}(r,z)$ are the fluid density and radial velocity at a cylindrical radial distance $r$ and at height $z$ from the equatorial plane of the disk. Now averaging in $z$ direction over disk height $H$
\begin{equation}
\frac{\partial \rho(r)}{\partial t}+\frac{1}{rH(r)}\frac{\partial }{\partial r}(\rho(r) v(r)rH(r))=0
\end{equation}
where $\rho(r)$ and $\rho(r)v(r)$ are the averaged fluid density and momentum respectively. The problem is now reduced in 1+1 dimension. Euler momentum equation is given by
\begin{equation}
\frac{\partial v}{\partial t}+v\frac{\partial v}{\partial r}=-\psi '(r)-\frac{1}{\rho}\frac{\partial p}{\partial r}+\frac{\lambda^2}{r^3}
\end{equation} 
where $\psi'(r)$ is the external field term and in this case it is gravitational force per unit mass of fluid exerted  by the accretor. $\lambda$ is the angular momentum of the fluid having small value.
Bernoulli's constant is given by
\begin{equation}
\zeta=\frac{1}{2}v^2+\int\frac{dp}{\rho}+\psi(r)+\frac{\lambda^2}{2r^2}
\end{equation}
Considering thin disk in vertical equilibrium and adiabatic flow, balancing pressure gradient force term and gravitational force term along z direction, vertical equilibrium condition is given(\cite{Nag_role_of_flow_geometry,Chaudhury2006})
\begin{equation}
H(r)=c_s(r)\sqrt{\frac{r}{\gamma\psi'}}
\end{equation}
As a consequence, continuity equation is given by
\begin{equation}
\partial_t(\rho^{\frac{\gamma+1}{2}})+\frac{\sqrt{\psi'}}{r^{\frac{3}{2}}}\partial_r\left(\frac{\rho^{\frac{\gamma+1}{2}}vr^{\frac{3}{2}}}{\sqrt{\psi'}}\right)
\end{equation}
$\gamma$ is specific heat ratio.
Introducing linear perturbation in the adiabatic flow, perturbation of Bernoulli's constant is given by
\begin{equation}
\zeta'(r,t)=v_0v'+\frac{c_{s0}^2\sigma}{\rho_0^{\frac{(\gamma+1)}{2}}}\delta(\rho^{\frac{(\gamma+1)}{2}})
\end{equation}
where $\sigma=\frac{2}{\gamma+1}$ and $\delta(\rho^{\frac{(\gamma+1)}{2}})$ is linear perturbation in $\rho^{\frac{(\gamma+1)}{2}}$. \\
Linear perturbation of Bernoulli's constant obeys massless scalar wave equation
\begin{equation}
\partial_{\mu}(f^{\mu\nu}(r)\partial_{\nu})\zeta'(r,t)=0
\end{equation}
where
\begin{equation}
f^{\mu\nu}(r)=\frac{\rho_0^{\frac{(\gamma+1)}{2}}r^{\frac{3}{2}}}{c_{s0}^2\sigma\sqrt{\psi'}}\begin{bmatrix}
-1  &~~~ -v_{0} \\
-v_{0}&~~~ \sigma c_{s0}^{2}-v_{0}^{2}
\end{bmatrix}
\end{equation}
$f_{\mu\nu}$ is taken as effective metric. Hence 2$\times$2 effective acoustic metric is given by
\begin{equation}
g_{\mu\nu}^{eff}(r)=\frac{\sqrt{\psi'}}{\rho_0^{\frac{(\gamma+1)}{2}}r^{\frac{3}{2}}}\begin{bmatrix}
 -(\sigma c_{s0}^{2}-v_{0}^{2}) &~~~ -v_{0}\\
-v_{0} &~~~ 1
\end{bmatrix}
\end{equation}
For isothermal flow one similarly gets acoustic metric like equation (34) where $\gamma$ is 1 and sound speed is a constant number.
\subsection{Constant height disk accretion}
In case of constant thickness model, $H$ is a constant. The linear perturbation of Bernoulli's constant is given by
\begin{equation}
\zeta'(r,t)=v_0v'+\frac{c_{s0}^2}{\rho_0}\rho'
\end{equation}
The linear perturbation of continuity equation is given by
\begin{equation}
\partial_t(\rho')+\frac{1}{rH}\partial_r(rH(\rho'v_0+\rho_0v'))=0
\end{equation}
where H is a non zero constant number.
The linear perturbation of momentum equation is given by
\begin{equation}
\partial_t(v')+\partial_r(\zeta')=0
\end{equation}
Now proceeding in the same way discussed earlier one gets equation of massless scalar field in curved space time background
\begin{equation}
\partial_\mu f^{\mu\nu}(r)\partial_\nu\zeta'=0
\end{equation}
where after taking inverse of $f^{\mu\nu}(r)$,$f_{\mu\nu}(r)$ can be taken as $2\times2$ effective metric as
\begin{equation}
f_{\mu\nu}=g_{\mu\nu}^{eff}(r)=\frac{1}{\rho_{0}rH}\begin{bmatrix}
 -(c_{s0}^{2}-v_{0}^{2}) &~~~ -v_{0}\\
-v_{0} &~~~ 1
\end{bmatrix}
\end{equation}
For conical disk model $H\propto r$. Just like the constant height disk model, linear perturbation in fluid does not have any influence on disk height. Hence the procedure of getting massless Klein Gordon equation is exactly same and the effective acoustic metric is exactly same as obtained in constant height disk model.
\section{Acoustic gravity in curved space-time background}
In the present work we consider the following metric for static space-time
\begin{equation}
ds^2=-g_{tt}dt^2+g_{rr}dr^2+g_{\theta\theta}d\theta^2+g_{\phi\phi}d\phi^2
\end{equation}
where the metric elements are functions of $r$ and can also be functions of $\theta$ and $\phi$.
We assume a perfect fluid with the energy-momentum tensor given by
\begin{equation}
T^{\mu \nu}=(\epsilon+p)v^\mu v^\nu+pg^{\mu \nu}
\end{equation}
with the velocity four-vector normalized as $v^\mu v_\mu=-1$ and $\epsilon$ is the internal energy per unit volume of the fluid.
The fluid is assumed to be ideal and so obeys equation of state for ideal gas. Also it is assumed to be under adiabatic condition i.e it obeys barotropic equation of state, i.e., $p=k\rho^\gamma$. The specific enthalpy of the fluid is given by
 \begin{equation}
h=\frac{\epsilon+p}{\rho}
\end{equation}
The speed of sound for adiabatic flow is given by 
\begin{equation}
c_s^2=\frac{\partial p}{\partial \epsilon}
\end{equation} which can be also written as(\cite{Bilic1999})
\begin{equation}
c_s^2=\frac{\rho}{h}\frac{\partial h}{\partial \rho}
\end{equation}
In our calculation of acoustic geometry we make use of two basic equations. First one is the continuity equation given by
\begin{equation}
\nabla_\mu (\rho v^\mu)=0
\end{equation}
and the second one is the irrotationality condition as the fluid is assumed to be irrotational. The condition is given by
\begin{equation}
\partial_\mu(hv_\nu)-\partial_\nu(hv_\mu)=0
\end{equation}
\subsection{Spherically symmetric radial flow}
In the first case in curved space-time background we derive the acoustic geometry for spherically symmetric flow. This implies that $v_\theta=v_\phi=0$ and all the derivatives with respect to $\theta$ and $\phi$ vanish.
Using $\mu=t$ and $\nu=r$ in the irrotationality condition equation given by equation (46) gives
\begin{equation}
\partial_t(hv_r)-\partial_r(hv_t)=0
\end{equation}
In stationary case where $\partial_t$ term vanishes the above equation implies $\partial_r(hv_t)=0$. So for stationary flow $\zeta=hv_t$ is a constant of the flow. This is called the specific energy for adiabatic flow or the Bernoulli's constant. 
The continuity equation given by equation (45) becomes
\begin{equation}
\frac{1}{\sqrt{-g}}\partial_t(\sqrt{-g}\rho v^t)+\frac{1}{\sqrt{-g}}\partial_r(\sqrt{-g}\rho v)=0
\end{equation}
where $v=v^r$ is the radial velocity. Using the normalization condition of the four-velocity given $v^\mu v_\mu=-1$, $v^t$ can be expressed as
\begin{equation}
 v^t=\sqrt{\frac{1+g_{rr}v^2}{g_{tt}}}
 \end{equation}  
 Now we linearly perturb the radial velocity, density and the Bernoulli's constant about their stationary values.
 \begin{equation}
 v(r,t)=v_0(r)+v'(r,t)
 \end{equation}
 \begin{equation}
 \rho(r,t)=\rho_0(r)+\rho'(r,t)
 \end{equation}
 and 
 \begin{equation}
 \zeta(r,t)=\zeta_0+\zeta'(r,t)
 \end{equation}
 
 Using these quantities we do linear perturbation of the continuity equation and the irrotaionality condition equation given by equation (48) and equation (47) respectively.
 
  Linear perturbation of the irrotationality condition equation gives the following equation
  \begin{equation}
  \partial_r\zeta'=g_{rr}h_0\partial_tv'+\frac{g_{rr}h_0v_0c_{s0}^2}{\rho_0}\partial_t\rho'
  \end{equation}
  where $h_0$ is the stationary or background value of the enthalpy $h$ and $c_{s0}^2=\frac{\rho_0}{h_0}\frac{\partial h}{\partial \rho}$
  again perturbing $\zeta=hv_t=-g_{tt}hv^t$ gives the equation 
  \begin{equation}
  \zeta'=-g_{tt}h_0\alpha v'-\frac{g_{tt}v_0^th_0c_{s0}^2}{\rho_0}\rho'
  \end{equation}
  where $\alpha=\frac{g_{rr}v_0}{g_{tt}v_0^t}$ and we have used the normalization condition of four-velocity to obtain $(v^t)'=\alpha v'$. Taking time derivative of the above equation gives 
  \begin{equation}
  \partial_t\zeta'=-g_{tt}h_0\alpha \partial_t v'-\frac{g_{tt}v_0^th_0c_{s0}^2}{\rho_0}\partial_t\rho'
  \end{equation}
  Using equation (53) and equation (55) we are  able to write $\partial_t v'$ and $\partial_t \rho'$ in terms of $\zeta'$ only. Thus we have
  \begin{equation}
  \partial_t v'=\frac{-1}{\Delta}[\frac{g_{rr}h_0v_0c_{s0}^2}{\rho_0}\partial_t \zeta'+\frac{g_{tt}v_0^th_0c_{s0}^2}{\rho_0}\partial_r \zeta']
  \end{equation}
  \begin{equation}
  \partial_t \rho'=\frac{1}{\Delta}[g_{rr}h_0\partial_t \zeta'+g_{tt}h_0\alpha \partial_r \zeta']
  \end{equation}
  where $\Delta=-\frac{g_{rr}h_0^2c_{s0}^2}{\rho_0v_0^t}$
  
  Linear perturbation of the continuity equation gives
  \begin{equation}
  \rho_0 \alpha \partial_t v'+v_0^t\partial_t \rho+\frac{1}{\sqrt{-g}}\partial_r(\sqrt{-g}\rho_0v'+\sqrt{-g}v_0\rho')=0
  \end{equation}
  taking the time derivative of the above equation gives
  \begin{equation}
  \partial_t(\sqrt{-g}\rho_0 \alpha \partial_t v')+\partial_t(\sqrt{-g}v_0^t\partial_t \rho)+\partial_r(\sqrt{-g}\rho_0\partial_tv')+\partial_r(\sqrt{-g}v_0\partial_t\rho')=0
  \end{equation}
  Substituting $\partial_t v'$ and $\partial_t \rho'$ in the above equation using equation (56) and equation (57) gives
  \begin{equation}
  \begin{split}
  \partial_t[\frac{\sqrt{-g}g_{rr}h_0}{\Delta v_0^t}\{\frac{g_{tt}(v_0^t)^2(1-c_{s0}^2)+c_{s0}^2}{g_{tt}}\}\partial_t \zeta']+\partial_t[\frac{\sqrt{-g}h_0g_{rr}v_0}{\Delta}\{1-c_{s0}^2\}\partial_r\zeta']\\
 \quad+\partial_r[\frac{\sqrt{-g}h_0g_{rr}v_0}{\Delta}\{1-c_{s0}^2\}\partial_t\zeta']+\partial_r[\frac{\sqrt{-g}g_{rr}h_0}{\Delta v_0^t}\{\frac{v_0^2g_{rr}(1-c_{s0}^2)-c_{s0}^2}{g_{rr}}\}\partial_r\zeta']=0
  \end{split}
  \end{equation}
  The above equation is of the form $\partial_\mu(f^{\mu\nu}\partial_\nu \zeta')=0$ with $f^{\mu\nu}$ given by after multiplying by -1
  
 \begin{equation}
f^{\mu\nu}=\frac{\sqrt{-g}\rho_0}{h_0}\begin{bmatrix}
    -g^{tt}+(v_{0}^t)^2(1-\frac{1}{c_{s0}^2}) & v_0v_0^t(1-\frac{1}{c_{s0}^2})\\\\
   v_0v_0^t( 1-\frac{1}{c_{s0}^2})&  g^{rr}+v_0^2(1-\frac{1}{c_{s0}^2})
    \end{bmatrix}
 \end{equation}
\subsection{Axially symmetric disk flow}
Three low angular momentum disk models (as discussed in previous sections) are considered for adiabatic flow. The normalization condition is given by
\begin{equation}
v_\mu v^\mu=-1
\end{equation}
The spherically symmetric diagonal metric of equation (40) is considered. We consider the dynamics only on the equatorial plane ($\theta=\frac{\pi}{2}$) plane of the disk. The accretion flow is irrotational, i.e., it obeys equation (46). The infalling fluid has a small azimuthal component of velocity, $v^\phi$. From equation (62)
\begin{equation}
(v^t)^2=\frac{1+g_{rr}(r)v^2+g_{\phi\phi}(r)(v^{\phi})^2}{g_{tt}(r)}
\end{equation}
Similarly, equation (47) gives Bernoulli's constant. Using equation equation (46) and axial symmetry of the flow
\begin{align*}
\partial_t(hv_\phi)=0&\\
\partial_r(hv_\phi)=0
\end{align*}
\begin{equation}
\Rightarrow hv_\phi=\rm constant=\ell
\end{equation}
$hv_\phi$ is called specific angular momentum and it is a constant number for non-stationary flow as well due to irrotationality and azimuthal symmetry. We assume that there is a stationary solution of the accretion($v_0(r), \rho_0(r), v^{\phi}_0(r), \zeta_0$) and linear perturbation is introduced.
 \begin{equation}
 v(r,t)=v_0(r)+v'(r,t)
 \end{equation}
 \begin{equation}
 \rho(r,t)=\rho_0(r)+\rho'(r,t)
 \end{equation}
 \begin{equation}
 v^\phi(r,t)=v^{\phi}_0(r)+v_\phi'(r,t)
 \end{equation} 
 and 
 \begin{equation}
 \zeta(r,t)=\zeta_0+\zeta'(r,t)
 \end{equation}
 The symbols carries usual meaning as before. The addition of linear perturbations do not make the accretion flow to violate irrotationlity, azimuthal symmetry (obvious from the above expressions). The accretion flow is still inviscid and adiabatic. From equation (64), linear perturbation term, $\ell'$ is given by
 \begin{equation}
 \ell'=0
 \end{equation}
 Using equation (44) and equation (69) we get
 \begin{equation}
 v_{\phi}'=-\frac{v^{\phi}_0 c_{s0}^2}{\rho_0}\rho'
 \end{equation}
 Using equation (63) and equation (70) we get
 \begin{equation}
 (v^t)'=\alpha_1(r) v'+\alpha_2(r)\rho'
 \end{equation}
 where 
\begin{align*}
&\alpha_1(r)=\frac{g_{rr}v_0}{g_{tt}v_{0}^t}\\
&\alpha_2(r)=-\frac{g_{\phi\phi}(v_0^\phi)^2c_{s0}^2}{g_{tt}v_0^t\rho_0}
\end{align*}
Irrotaionality condition gives
\begin{equation}
\partial_r\zeta'=f_1(r)\partial_t\rho'-f_2(r)\partial_tv'
\end{equation} 
where
\begin{align*}
& f_1(r)=\frac{g_{rr}v_0h_0c_{s0}^2}{\rho_0}\\
& f_2(r)=-g_{rr}h_0
\end{align*}
Using equation (71) and expression of $\zeta(=-hg_{tt}v^t)$
\begin{equation}
\partial_t\zeta'=-f_3(r)\partial_t\rho'+f_4(r)\partial_tv'
\end{equation}
where
\begin{align*}
& f_3(r)=\frac{g_{tt}v_0^th_0c_{s0}^2}{\rho_0}+g_{tt}h_0\alpha_2\\
& f_4(r)=-g_{tt}h_0\alpha_1
\end{align*}
From equation (72) and (73) we get
\begin{equation}
\partial_t\rho'=g_4(r)\partial_r\zeta' + g_2(r)\partial_t\zeta'
\end{equation}
\begin{equation}
\partial_tv'=g_3(r)\partial_r\zeta' + g_1(r)\partial_t\zeta'
\end{equation}
where
\begin{align*}
& g_i=\frac{f_i}{\Delta} ~~ \text{where}~  	i\in {\rm\mathbb{N}}~\text{and}~ i=1~\rm to~ 4~ where ~\mathbb{N}~is~the~set~of~natural~numbers. \\
& ~\text{and}~\Delta=f_1f_4-f_2f_3=\frac{g_{rr}h_0^2c_{s0}^2}{\rho_0v_0^t}
\end{align*}
Until and unless we don't have the expression of disk height, we can not use the continuity equation. In the next section several sub-Keplarian (discussed before) disk models are considered.
\subsubsection{\bf Vertical equilibrium disk model}
The expression of $H(r)$\footnote[1]{$H(r)$ is not the flow thickness of the disk, this is a dimensionless quantity which appears in continuity equation after averaging in $\theta$ direction.} satisfying vertical equilibrium condition is given by(\cite{Abramowicz1996ap,deepika_ax_schwarzchild})
\begin{align*}
H(r)^2v_\phi^2F(r)=\frac{p}{\rho}
\end{align*}
The linear perturbation of $H$ is $H'$ and stationry solution of $H$ is $H_0(r)$. Using barotropic equation and equation (70)
\begin{equation}
\frac{\partial_t H'}{H_0}=\frac{\beta}{\rho_0}\partial_t\rho'
\end{equation}
where $\beta=c_{s0}^2+\frac{\gamma-1}{2}$.\\
Continuity equation is given by 
\begin{equation}
\frac{1}{\sqrt{-g}}\partial_t(\sqrt{-g}\rho v^tH)+\frac{1}{\sqrt{-g}}\partial_r(\sqrt{-g}\rho vH)=0
\end{equation}
Introducing linear perturbation in the fluid and using equation (71) and (76), one gets after partially differentiating equation (77),
\begin{equation}
(\partial_tF_1\partial_t)\rho'+(\partial_tF_2\partial_t)v'+(\partial_rF_3\partial_t)\rho'+(\partial_rF_4\partial_t)v'=0
\end{equation}
where each $F_i$ $(i\in \mathbb{N}, i=1~\rm to~ 4)$ is function of $r$, the expressions are given below
\begin{align*}
& F_1(r)=\sqrt{-g}(H_0v_0^t(1+\beta)+\alpha_2\rho_0H_0)\\
& F_2(r)=\sqrt{-g}\rho_0H_0\alpha_1\\
& F_3(r)=\sqrt{-g}H_0v_0(1+\beta)\\
& F_4(r)=\sqrt{-g}\rho_0H_0
\end{align*}
Using equation (74), (75) and (78) we get 
\begin{equation}
\partial_\mu (f^{\mu\nu}(r)\partial_\nu)\zeta'=0
\end{equation}
where $\mu, \nu$ indices run over $t$ and $r$. 2$\times$2 matrix, $f^{\mu\nu}$ is given by
   \begin{equation}
  f^{\mu\nu}=\frac{\sqrt{-g}H\rho_0}{h_0}\begin{bmatrix}
    -g^{tt}+(v_{0}^t)^2(1-\frac{1+\beta}{c_{s0}^2}) & v_0v_0^t(1-\frac{1+\beta}{c_{s0}^2})\\\\
   v_0v_0^t( 1-\frac{1+\beta}{c_{s0}^2})&  g^{rr}+v_0^2(1-\frac{1+\beta}{c_{s0}^2})
    \end{bmatrix}
   \end{equation}
 \subsubsection{\bf Constant height disk model}
 For constant height disk model $H\propto \frac{1}{r}$. $H$ does not change when linear perturbations are introduced in the fluid velocity and density. Now using continuity equation (77) and introducing linear perturbations, one gets after partially differentiating with $t$
 \begin{equation}
 (\partial_tF_1\partial_t)\rho'+(\partial_tF_2\partial_t)v'+(\partial_rF_3\partial_t)\rho'+(\partial_rF_4\partial_t)v'=0
 \end{equation}
 where each $F_i$ $(i\in \mathbb{N}, i=1~\rm to~ 4)$ is function of $r$, the expressions are given below
 \begin{align*}
 & F_1(r)=\sqrt{-g}(Hv_0^t+\alpha_2\rho_0H)\\
 & F_2(r)=\sqrt{-g}\rho_0H\alpha_1\\
 & F_3(r)=\sqrt{-g}Hv_0\\
 & F_4(r)=\sqrt{-g}\rho_0H
 \end{align*}
 Using equation (74), (75) and (81) we get 
 \begin{equation}
 \partial_\mu (f^{\mu\nu}(r)\partial_\nu)\zeta'=0
 \end{equation}
 where $\mu, \nu$ indices run over $t$ and $r$. 2$\times$2 matrix, $f^{\mu\nu}$ is given by
   \begin{equation}
  f^{\mu\nu}=\frac{\sqrt{-g}H\rho_0}{h_0}\begin{bmatrix}
    -g^{tt}+(v_{0}^t)^2(1-\frac{1}{c_{s0}^2}) & v_0v_0^t(1-\frac{1}{c_{s0}^2})\\\\
   v_0v_0^t( 1-\frac{1}{c_{s0}^2})&  g^{rr}+v_0^2(1-\frac{1}{c_{s0}^2})
    \end{bmatrix}
   \end{equation}
 \subsubsection{\bf Conical disk model}
For conical disk model, H is a constant number. $ H $ does not change when linear perturbations are introduced in the fluid velocity and density because $ H $ does not depend on those quantities. Now using continuity equation (77) and introducing linear perturbations, one gets after partially differentiating with $ t $
  \begin{equation}
  (\partial_tF_1\partial_t)\rho'+(\partial_tF_2\partial_t)v'+(\partial_rF_3\partial_t)\rho'+(\partial_rF_4\partial_t)v'=0
  \end{equation}
 where each $F_i$ $(i\in \mathbb{N}, i=1~\rm to~ 4)$ is function of $r$, the expressions are given below
  \begin{align*}
  & F_1(r)=\sqrt{-g}(Hv_0^t+\alpha_2\rho_0H)\\
  & F_2(r)=\sqrt{-g}\rho_0H\alpha_1\\
  & F_3(r)=\sqrt{-g}Hv_0\\
  & F_4(r)=\sqrt{-g}\rho_0H
  \end{align*}
  Using equation (74), (75) and (84) we get 
  \begin{equation}
  \partial_\mu (f^{\mu\nu}(r)\partial_\nu)\zeta'=0
  \end{equation}
where $\mu, \nu$ indices run over $t$ and $r$. 2$\times$2 matrix, $f^{\mu\nu}$ is given by
   \begin{equation}
  f^{\mu\nu}=\frac{\sqrt{-g}H\rho_0}{h_0}\begin{bmatrix}
    -g^{tt}+(v_{0}^t)^2(1-\frac{1}{c_{s0}^2}) & v_0v_0^t(1-\frac{1}{c_{s0}^2})\\\\
   v_0v_0^t( 1-\frac{1}{c_{s0}^2})&  g^{rr}+v_0^2(1-\frac{1}{c_{s0}^2})
    \end{bmatrix}
   \end{equation}
 One trivial observation is that for constant height disk model and conical disk model as the disk height is not disturbed due to linear perturbations in the fluid, hence putting $\beta$ to be zero in the matrix (80) one can obtain $f^{\mu\nu}$ for these models.
 \section{Astrophysical significance}
The linear perturbation of Bernoulli's constant satisfies a wave equation in the case of accretion onto a black hole or a massive body. The generic equation is
\begin{equation}
\partial_\mu (f^{\mu\nu}(X)\partial_\nu)\zeta'=0
\end{equation}
where $\zeta'=\zeta'(X,t)$ and $X$ is spherical polar radial coordinate for spherically symmetric accretion or cylindrical polar radial coordinate for axisymmetric accretion. The above equation is true for both general relativistic framework as well as for Newtonian-gravity framework. The original fluid equations, i.e., continuity equation, Euler equation, are time dependent partial differential equations. Here our approach is perturbative, i.e., there is a existing steady state solution for astrophysical accretion problem and over that we are introducing linear perturbation of the fluid quantities. Now the solution of $\zeta'$ of the equations could tell us whether the steady state solutions are stable under small perturbations or not. As the linear perturbation of density, velocity are all related to $\zeta'$, hence if we could find certain conditions under which $\zeta'$ grows in time then the other related quantities would grow in time, the perturbation could not be small at all time, the steady solution of the quantities would not be stable in that case and in that case a full numerical approach where partial time derivatives in the fluid equations are taken care of would give a more accurate result in stead of steady state solution.

To find $\zeta'$, we use the same approach done by Jacobus A. Petterson et al (\cite{Pson}). We take the form of the wave to be as
\begin{equation}
\zeta'(X,t)=P_{\omega}(X)e^{i\omega t}
\end{equation} 
Hence from equation (87), we find
\begin{equation}
\omega^2 P_\omega(X)f^{tt} + i\omega [\partial_X\left(P_\omega(X)f^{Xt}\right) + f^{tX}\partial_XP_\omega(X)]-[\partial_X \left (f^{XX} \partial_{X}P_\omega(X)\right)] = 0.
\end{equation}
\subsection{Standing wave analysis}
Standing wave means that there will be nodes and antinodes. There will exist a standing wave across $X$ direction if and only if there are at least two different $X$s in space, called nodes, where the amplitude of the $\zeta'(X,t)$ is zero for all time where unlike $\theta,~\phi$; $X$ is a noncompact dimension. Therefore, there exist at least two radii $X_1,~X_2;~X_1\neq X_2$ such that $P_\omega(X_1)=P_\omega(X_2)=0$. Hence $\zeta'$ is zero at these two points for all time. Standing wave is produced when two waves moving in opposite direction superpose with each other in space. When a wave moving along a particular direction face an obstacle another wave moving along the opposite direction is produced due to the reflection from that obstacle and in time superposition of these two waves produce standing wave confined between two points or radii in space. The outer radius, say $X_2$ may be a very large radius, for example the boundary of the accreting cloud surrounding the star. In case of accretion on to a black hole, there is no solid surface to produce a standing wave by reflection, i.e., in the supersonic region of flow, nowhere $\zeta'$ is zero, i.e., there is no inner radius to confine the wave between two radii which is required to produce a standing wave. In the case of accretion on to a compact object like neutron star, on the surface of the star, the accreting fluid collides, hence the fluid quantities undergo a discontinuity, as a result of this, a shock wall is formed according to Rankine-Hugoniot relations around the star where the pre shock supersonic inflow becomes post shock subsonic flow. As in our analysis we are not considering any discontinuity in the fluid equations. We restrict this analysis for subsonic flows.

Integrating the equation (89) between $X_1$ and $X_2$ and imposing the condition of vanishing amplitude of $\zeta'$ at the two radii, we get
\begin{equation}
\omega^2=-\frac{C}{A}
\end{equation}
where
\begin{align*}
& A=\int\limits_{X_1}^{X_2}(P_\omega)^2f^{tt}dX\\
& C=\int\limits_{X_1}^{X_2}(\partial_XP_\omega)^2f^{XX}dX
\end{align*}
For linear perturbation of mass accretion rate, $\omega^2$ happens to be positive(\cite{Pson}\cite{deepika-sph2015}\cite{deepika_ax_schwarzchild}). As $f^{\mu\nu}$ for linear perturbation of Bernoulli's constant and $f^{\mu\nu}$ for the linear perturbation of mass accretion rate only differ by a conformal factor, hence the sign of the conformal factor cancels out in the numerator and denominator. Hence the conclusion is same for both the perturbations. Hence $\zeta'$ is of oscillatory kind, it does not blow up with time.

We introduced linear perturbation in the fluid medium by introducing linear perturbation in density and velocity of the medium. We see that linear perturbation, $\zeta'$ is the linear combination of linear perturbation of density and linear perturbation of radial fluid velocity, it has the following generic structure for all the cases.
\begin{equation}
\zeta'(X,t)=f_{\zeta\rho}(X)\rho'(X,t)+f_{\zeta v}(X)v'(X,t)
\end{equation}
where $f_{\zeta\rho}(X)$ and $f_{\zeta v}(X)$ are the functions of the radial coordinate $X$.\\
Linear perturbation of mas accretion rate, $\lambda'$ also has the similar structure.
\begin{equation}
\lambda'(X,t)=f_{\lambda\rho}(X)\rho'(X,t)+f_{\lambda v}(X)v'(X,t)
\end{equation} 
In other words the above two equations can be expressed as a matrix equation
\begin{equation}
\begin{pmatrix} \zeta' \\  \lambda' \end{pmatrix}=
\begin{pmatrix} f_{\zeta\rho} & f_{\zeta v}\\ f_{\lambda\rho} & f_{\lambda v} \end{pmatrix}
\begin{pmatrix} \rho' \\ v' \end{pmatrix} =
\hat{f} \begin{pmatrix} \rho' \\ v' \end{pmatrix}
\end{equation}
The expressions of $f'$ and $\zeta'$ for all the cases show that $det(\hat{f})$ is nonzero in general for subsonic flows, i.e., $\hat{f}$ is a non-singular matrix. One easy way to see this is that $f_{\zeta\rho}(X)$ contains $c_{s0}^2$ but $f_{\zeta v}(X)$, $f_{\lambda\rho}(X)$, $f_{\lambda v}(X)$ do not contain $c_{s0}^2$. For subsonic flow, the right hand side of the equation (91) and the right hand hand side of the equation (92) do not differ just by a conformal factor and this implies non-singularity of the matrix $\hat{f}$. Hence both $\rho'$ and $v'$ can be expressed as a linear combination of $\zeta'$ and $\lambda'$. Now considering the physical situation, $X_1$, $X_2$ are same for both $\zeta'$ and $\lambda'$. Hence at these two radii, not only $\zeta'$ and $\lambda'$ vanish but also $\rho'$ and $v'$ vanish. In the next section we have introduced some thermodynamic quantities like entropy, temperature. Since $\rho'$ is zero at these two radii and the linear perturbations of the thermodynamic quantities like enthalpy, entropy are proportional to $\rho'$, the linear perturbation of all thermodynamic quantities are zero at these two radii. Similarly the linear perturbation of dynamical quantities like kinetic energy per unit mass are zero at these two radii because linear perturbation of the dynamical quantities are proportional to $v'$ \footnote{there is an exception, in the case of axially symmetric disk accretion we have seen that $v_{\phi}'\propto\rho'$}.

As the standing wave analysis, valid for subsonic flows, show that both $\zeta'$ and $\lambda'$ do not blow in time, the linear perturbation of all the dynamical quantities and the thermodynamic quantities do not blow in time too. Therefore, the steady state solution for subsonic flows are stable under standing wave perturbations.

The steady state solutions are governed by the integrals of motions, i.e., Bernoulli's constant, mass accretion rate and specific angular momentum (for axisymmetric disk accretion). For the disk accretion models which we have considered, the linear perturbation of specific angular momentum happens to be zero due to symmetry. The non-trivial linear perturbations are the perturbation of the accretion rate and the perturbation of Bernoulli's constant.  For time dependent case in the linear perturbation method, the time dependent solutions are also governed by the linear perturbation of these two integrals of motion.
\subsection{Travelling wave analysis}
We study high frequency travelling wave. We assume the wavelength of the wave to be smaller than the relevant smallest length scale of the problem. W assume the travelling wave to be of the following form. 
\begin{equation}
P_{\omega}(X)=exp\left[\sum_{n=-1}^{n=\infty}\frac{K_n(X)}{\omega^n}\right]
\end{equation}
Using equation (88) and equating the coefficients of $\omega$ and $\omega^2$, we get
\begin{equation}
K_{-1}(X)=i\int^{X}\frac{f^{Xt}\pm\sqrt{(f^{Xt})^2-f^{tt}f^{XX}}}{f^{XX}}dX
\end{equation} 
\begin{equation}
K_{0}(X)=-\frac{1}{2}ln\left(\sqrt{(f^{Xt})^2-f^{tt}f^{XX}}\right)
\end{equation} 
From the expression, it is obvious that $K_{-1}(X)$ is a purely imaginary quantity. For consistency in the solution, the following relation must hold.
\begin{equation}
\omega K_{-1}(X)>>K_0(X)>>...
\end{equation}
\subsubsection{Nonrelativistic framework}
In Newtonian-gravity framework, for all the geometries (spherically symmetric accretion and disk geometries) discussed earlier, $K_{-1}(X)$ and $K_0(X)$ have the following general structure.
\begin{equation}
K_{-1}(X)=i\int_{}^{X}\frac{1}{v_0(X)\pm\sqrt{\sigma}c_{s0}(X)}dX
\end{equation}
and 
\begin{equation}
K_0(X)={\rm constant}+\frac{1}{2}ln\left(v_0(X)c_{s0}(X)\right)
\end{equation}
where $\sigma=\frac{2}{\gamma+1}$ for vertical equilibrium disk accretion and $\sigma=1$ for spherically symmetric accretion and other disk geometries. The expressions are very similar to the case of linear perturbation of mas accretion rate. At large radii, equation (97) is true due to the virtue of high frequency approximation. The expressions of $K_{-1}(X)$ demonstrates that near the critical point, for the wave travelling upstream of motion, $\frac{\partial K_{-1}(X)}{\partial X}$ is very large near event horizon. The condition (97) is well satisfied there. The conclusions are same for both the cases, i.e., the case of linear perturbation of mas accretion rate and the case of linear perturbation of Bernoulli's constant.
\subsubsection{General relativistic framework} 
The expression of $K_0$ and $K_{-1}$ is given by
\begin{equation}
K_0(X)={\rm constant}+\frac{1}{2}ln\left(\frac{v_0c_{s0}\sqrt{g_{tt}g_{XX}}}{v_0^t}\Omega(X) \right)
\end{equation}
and 
\begin{equation}
K_{-1}(X)=i\int^{X}dX\frac{\left(v_0v_0^t(1-\frac{\sigma}{c_{s0}^2})\pm\frac{1}{\sqrt{g_{tt}g_{XX}}c_{s0}}\Omega(X)\right)}{g^{XX}+v_0^2(1-\frac{\sigma}{c_{s0}^2})}
\end{equation}
where 
\begin{align*}
&\sigma=1+\beta~~~{\rm for~vertical~equilibrium~disk~model}\\
&\sigma=1~~~{\rm for~spherically~symmetric~radial~accretion~and~other~disk~models}\\
&\Omega(X)=1~~~{\rm for~spherically~symmetric~radial~flow}\\
&\Omega(X)=\sqrt{\sigma+g_{\phi\phi}(v_0^\phi)^2(\sigma-c_{s0}^2)}~~~{\rm for~disk~accretion}
\end{align*}
$\frac{\partial K_{-1}(X)}{\partial X}$ is very large near the critical point as the denominator approaches zero there. Hence condition (97) is well satisfied near sonic horizon. For very large $X$, $K_{0}(X),~K_{-1}(X)$ are close to non-relativistic values, the criteria (97), is true at large $X$ due to the virtue of high $\omega$.  
 \section{Concluding remarks}
 Linear perturbation of several quantities obey massless scalar field equation in acoustic space time background. Unruh(\cite{Unruh}) first shown that linear perturbation of velocity potential obeys massless scalar field equation in curved space time background. In these papers(\cite{deepika-sph2015,deepika_ax_schwarzchild}) , it is explicitly shown that linear perturbation of mass accretion rate also gives acoustic metric. In our paper, we've shown that Bernoulli's constant also produces analogue gravity.
 
 For non-general relativistic background flow of adiabatic fluid, the Bernoulli's constant can be expressed as an additive term of various energy contribution to the total energy of the system. If one is interested to learn how the various sources of energy of the system, i.e., gravitational, mechanical, thermal and rotational, gets perturbed individually, the perturbation scheme of the Bernoulli's constant will be of great help to understand such physics and related issues. For instance, one can directly connect the Bernoulli's constant to some dynamical and thermodynamic energy quantities. One can define specific total energy(\cite{clarke}) as 
 \begin{align*}
 E=(\frac{1}{2}u^2+\xi+V_{ext})
 \end{align*}
 where $\xi$ is specific internal energy. After a little manipulation, using equation (10), i.e., for non relativistic case, one can find a set of equations relating some dynamical energy quantities and thermodynamic quantities for adiabatic flow that 
 \begin{align*}
& \partial_{\mu}(f^{\mu\nu}(\vec{x})\partial_{\nu})E'(\vec{x},t)+ \frac{\gamma-1}{\gamma}\partial_{\mu}(f^{\mu\nu}(\vec{x})\partial_{\nu})h'=0\\
&\partial_{\mu}(f^{\mu\nu}(\vec{x})\partial_{\nu})E'(\vec{x},t)+(\gamma-1)\partial_{\mu}(f^{\mu\nu}(\vec{x})\partial_{\nu})\xi'(\vec{x},t)=0\\
&\partial_{\mu}(f^{\mu\nu}(\vec{x})\partial_{\nu})E'(\vec{x},t)+(\gamma-1)(\partial_{\mu}(f^{\mu\nu}(\vec{x})\partial_{\nu})F'+s_0 \partial_{\mu}(f^{\mu\nu}(\vec{x})\partial_{\nu})T'=0\\
&\partial_{\mu}(f^{\mu\nu}(\vec{x})\partial_{\nu})E'(\vec{x},t)+\frac{\gamma-1}{\gamma}\partial_{\mu}(f^{\mu\nu}(\vec{x})\partial_{\nu})G'(\vec{x},t)+\frac{\gamma -1}{\gamma}s_0 \partial_{\mu}(f^{\mu\nu}(\vec{x})\partial_{\nu})T'=0
 \end{align*}
where $E'$, $h'$, $G'$, $F'$ and $\xi'$ are linear perturbation in $E$, specific enthalpy, specific Gibbs free energy, specific Helmholtz's free energy and specific internal energy. $s_0$ is constant entropy value for adiabatic case, i.e., $s_0=ln(\frac{p_0}{\rho_0^\gamma})$. 

For isothermal case we find that
\begin{align*}
& \partial_{\mu}(f^{\mu\nu}(\vec{x})\partial_{\nu})E'(\vec{x},t)+\partial_{\mu}(f^{\mu\nu}(\vec{x})\partial_{\nu})G'(\vec{x},t)=0\\
&\partial_{\mu}(f^{\mu\nu}(\vec{x})\partial_{\nu})E'(\vec{x},t)-\partial_{\mu}(f^{\mu\nu}(\vec{x})\partial_{\nu})F'(\vec{x},t)=0
\end{align*}
For General Relativistic case and for adiabatic flow,
$\because h=g-Ts$ and $\zeta=hv_t$, defining $\alpha'=(gv_t)'$ and $\beta'=(Tv_t)'$ 
\begin{equation*}
\partial_{\mu}(f^{\mu\nu}(\vec{r})\partial_{\nu})\alpha'(\vec{r},t)+s_0\partial_{\mu}(f^{\mu\nu}(\vec{r})\partial_{\nu})\beta'(\vec{r},t)=0
\end{equation*}
Clearly not all energy quantities satisfy the differential equation satisfied by linear perturbation of Bernoulli's constant but if one can perturb only the thermodynamic quantities without perturbing the dynamical quantities, then one can find again acoustic geometry and do analogue gravity on that.

 We introduce here another important finding of related interest.
 Linear perturbation of any algebraic function of $\zeta$ obeys massless scalar field equation in acoustic space time if the function and it's first derivative with respect to $\zeta$  exist at the background value, i.e., at $\zeta_0$. The wave equation satisfied by linear perturbation of the function is exactly the same as the wave equation obeyed by linear perturbation of $\zeta$. Same argument holds for the linear perturbation of mass accretion rate as well.
 To prove this, let's consider an algebraic function of $\zeta$, $F(\zeta)$. 
 We have 
 \begin{equation}
 \partial_\mu f^{\mu\nu}(\vec x)\partial_\nu\zeta'=0
 \end{equation}
 where $\zeta$ is perturbed linearly as
 \begin{align*}
 \zeta(\vec x,t)=\zeta_0(\vec x)+\zeta'(\vec x,t)
 \end{align*}
 \begin{align*}
 &\Rightarrow F(\zeta(\vec x, t))=F(\zeta_0(\vec x)+\zeta'(\vec x,t))\\
 &=F(\zeta_0+\zeta'(\vec x,t))\\
 &=F(\zeta_0)+\left(\frac{dF}{d\zeta}\right)_{\zeta_{0}}\zeta'(\vec x,t)\\
 &=F_0+F'
 \end{align*}
 $\zeta_0$, $\left(\frac{dF}{d\zeta}\right)_{\zeta_{0}}$ are constant numbers because $F(\zeta)$ and it's first derivative exist at $\zeta_0$ and $\zeta_0$ is a constant of motion. Linear perturbation of $F(\zeta)$ is a constant multiple of $\zeta'$. Hence linear perturbation of $F(\zeta)$ obeys exactly same massless scalar field equation as obeyed by $\zeta'$.
 
 This is another advantage of constants of motion, $\zeta$ and $f$, over the velocity potential $\psi$, is that one can construct infinitely many quantities with any of constants of motion, $\zeta$ or f, whose linear perturbation obeys massless scalar field equation in curved  space time.
 
 Similar argument holds for the case of mass accretion rate in 1+1 dimension as well.
 In this fashion one can construct two disjoint sets of algebraic functions from two independent constants of motion $\zeta$ and $f$ respectively. At this point, we thus argue that the linear perturbation of any quantity of fluid motion if obeys a massless scalar field equation, that equation will be same as wave equation satisfied by either of $\zeta'$ or $f'$.\\
In summary, one can also proceed with velocity potential in the analysis instead of Bernoulli's constant. here we have started by analysing the linear perturbation of Bernoulli's constant instead of velocity potential. This is a independent approach to the same problem of analogue gravity. There is so far no work, of course to the best of our knowledge, analysing linear perturbation of velocity potential in the case of vertical equilibrium disk model for both relativistic and non relativistic case; we find that for vertical equilibrium disk model (not shown for brevity) 
\begin{equation}
 \partial_\mu f^{\mu\nu}(\vec x)\partial_\nu\psi'=0
\end{equation}
where $\psi'$ is linear perturbation of velocity potential, $\psi$. One can use the relation $\partial_t\psi'=\zeta'$ to find 
\begin{equation}
 \partial_\mu f^{\mu\nu}(\vec x)\partial_\nu\zeta'=0.
\end{equation}
This is clearly an alternative approach.
\section{Acknowledgement}
The author would like to thank the anonymous referee for providing useful comments and suggestions. 

\section*{References}
\bibliographystyle{elsarticle-harv} 
\bibliography{reference_arif}

\end{document}